\documentclass[twocolumn,showpacs,preprintnumbers,amsmath,amssymb]{revtex4}
\usepackage{graphicx}
\usepackage{dcolumn}
\usepackage{bm}
\usepackage[latin1]{inputenc}
\usepackage{color}
\newcommand{\comment}[1]{}
\newcommand\etal{\mbox{\textit{et al.}}}

\begin{document}
\setlength{\unitlength}{0.7\textwidth} \preprint{}

\title{Origin of Lagrangian intermittency in drift-wave turbulence}

\author{B. Kadoch{$^1$}, W.J.T. Bos{$^2$} and K. Schneider{$^1$}}

\affiliation{$^1$ M2P2, CNRS UMR 6181 \& CMI, Ecole Centrale de
  Marseille, Universit\'es d'Aix-Marseille, Marseille, France}
\affiliation{$^2$ LMFA, CNRS UMR 5509, Ecole Centrale de Lyon -\\
Universit\'e de Lyon, Ecully, France}

\begin{abstract}
The Lagrangian velocity statistics of dissipative drift-wave
turbulence are investigated. 
For large values of the adiabaticity (or small collisionality), the probability density function of the Lagrangian acceleration shows exponential tails, as opposed to the stretched exponential or algebraic tails, generally observed for the highly intermittent acceleration of Navier-Stokes turbulence. This exponential distribution is shown to be a robust feature independent of the Reynolds number. For small adiabaticity, algebraic tails are observed, suggesting the strong influence of point-vortex-like dynamics on the acceleration. A causal connection is found between the shape of the  probability density function and the auto-correlation of the norm of the acceleration.
\end{abstract}


\maketitle

Turbulence is one of the main actors in degrading the confinement
quality of magnetically confined fusion plasmas. This so-called
micro-turbulence in the edge of plasma fusion devices, such as
tokamaks is commonly admitted to be of electrostatic nature
\cite{Scott2002,Horton1990}. A typical instability leading to this
turbulent motion is the drift-wave instability, present in plasmas
with a strong magnetic field and a temperature or pressure gradient.
Turbulence leads to an enhanced diffusivity and its average influence
can be characterized by transport coefficients which represent the
mean influence of turbulent motion as an enhanced fluid property
\cite{Boussinesq1877}. Reviews on the  use of transport coefficients
in fusion devices are given in
[\onlinecite{Garbet2001,Garbet2006}]. Transport coefficients allow to
describe the mean transport on the level of second order moments such
as the variance of the impurity density, kinetic energy and
fluxes. The spatial and temporal fluctuations around these variances
are however not described by such an approach, since they are directly
related to fourth-order moments. These fourth-order moments will give
a rough description of the intermittent properties of the turbulence:
{is the transport bursty, corresponding to non Gaussian fluctuations
  or diffusive so that it could be modeled by a Gaussian process?} Indeed, if the turbulent transport is dominated by rare but strong events, the impact on the confinement quality will be different from the case where a {Gaussian process} governs the transport. In three-dimensional fluid turbulence it is now well established that the velocity displays near Gaussian statistics but that the velocity gradients and acceleration are characterized by probability density functions (PDFs) with strongly non-Gaussian tails \cite{Yeung1989,Toschi2009}. In two-dimensions it was shown that Lagrangian statistics can be strongly non-Gaussian even when the Eulerian statistics are perfectly Gaussian \cite{Kamps2008}. The present investigation is dedicated to the characterization of Lagrangian intermittency in the close-to-two-dimensional dynamics of electrostatic plasma turbulence. 

Intermittency can be investigated through the statistical properties of velocity increments $\delta \bm u$, which can be defined both in an Eulerian and in a Lagrangian reference frame. Lagrangian velocity increments are defined as $\delta \bm u(t,\tau)=\bm u(t+\tau)- \bm u(t)$, where $\bm u(t)$ is the Lagrangian velocity, \emph{i.e.} the velocity of a passive tracer monitored on its trajectory as a function of time. When the shape of the PDF of the velocity increments varies as a function of $\tau$, the statistics are usually said to be intermittent, even though this definition can be criticized \cite{Bos2010-1}. At smallest $\tau$ the PDFs approach the shape of the acceleration PDF, which is generally non-Gaussian in turbulent flows. 

The study of the Lagrangian dynamics of fluid turbulence is now possible in controlled turbulence experiments in which small solid tracer particles are followed in the flow (\emph{e.g.} \cite{Ott2000,Mordant2001,Voth2002}) and numerical simulations of the Navier-Stokes equations \cite{Yeung1989}. Whereas the experimental tracing of particles in fusion reactors introduces problems related to the extreme conditions in controlled fusion, tracing of particles in numerical simulations of drift-wave turbulence is perfectly possible. In a recent study [\onlinecite{Bos2010-1}], we presented detailed results on the Lagrangian statistics obtained in simulations of drift-wave turbulence, within the context of the Hasegawa-Wakatani model \cite{Hasegawa1983,Wakatani1984}. In the present letter we will focus on the non-Gaussianity of the acceleration statistics. In particular we will investigate the influence of the Reynolds number and the collisionality on the statistics and we will propose explanations for the observed behavior.

The Hasegawa-Wakatani model can be derived from the Braginskii two-fluid description \cite{Braginskii1965}, considering an ion-fluid and an electron-fluid in the presence of a fixed magnetic field, assuming isothermal inertia-less electrons and cold ions. {For details on the derivation of the 2D slab version of Hasegawa-Wakatani equation, we refer  e.g. to \cite{Horton1996}.} The model assumptions yield eventually a closed set of equations, describing the vorticity $\omega=\nabla^2 \phi$ of the $E\times B$ motion (with $\phi$ the electrostatic potential) and the advection of the plasma density fluctuations $n$: 
\begin{eqnarray}
\left(\frac{\partial}{\partial{t}}- \nu\nabla^2\right)\nabla^2 \phi=
\left[\nabla^2 \phi,\phi\right]
+c(\phi-n),\label{hw1}\\
\left(\frac{\partial}{\partial{t}}- D\nabla^2\right)n=\left[n,\phi\right]-\bm u \cdot \nabla \ln(\left<n\right>) +c(\phi-n),\label{hw2}
\end{eqnarray}
in which all quantities are suitably normalized as in [\onlinecite{Bos2008-1}]. The model-equations closely ressemble the two-dimensional Navier-Stokes equations combined with the advection equation for a scalar $n$, representing here the fluctuations of the plasma density around a mean profile. Small scale damping is introduced through the Laplacians, with $\nu$ and $D$ denoting viscosity and diffusivity, respectively. Nonlinearities are written as Poisson brackets {$[a,b]=\frac{\partial a}{\partial x}\frac{\partial b}{\partial y}-\frac{\partial a}{\partial y}\frac{\partial b}{\partial x}$}. The source term in the above equation is the mean plasma density profile $\left<n\right>$, which is assumed to be exponentially decaying in the $x$-direction and homogeneous in the $y$-direction, so that equation (\ref{hw2}) reduces to the advection of a scalar fluctuation with respect to an imposed uniform mean scalar gradient. The electrostatic potential $\phi$ plays for the $E\times B$ velocity the role of a stream-function, $\bm u=\nabla_\perp \phi$, \emph{i.e.}, $u_x=-\partial \phi/\partial y$ and $u_y=\partial \phi/\partial x$.  The Lagrangian acceleration of tracer particles, advected by the $E\times B$ velocity is then
\begin{equation}\label{eqaL}
\bm a_L=\frac{\partial \nabla_\perp \phi}{\partial t}+\left[\phi,\nabla_\perp\phi\right] =-\nabla p+\nu\nabla^2\bm u- \frac{\nabla_\perp}{\nabla^2} \left[c(n-\phi)\right],
\end{equation}
where $p$ is the pressure. The adiabaticity $c$ is given by
\begin{equation}
c=\frac{T_e k_{z}^2}{e^2n_0\eta\omega_{ci}},
\end{equation}
with $T_e$ the electron-temperature, $k_{z}$ the effective parallel wavenumber, $e$ the electron charge,   $n_0$ the reference plasma density, $\eta$ the electron resistivity and $\omega_{ci}$ the ion-gyro-frequency. The adiabaticity is therefore determined by the electron resistivity, which is strongest in the edge of fusion devices, where the temperature drops. The proportionality to $k_z^2$, the square of the dominant wavenumber in the parallel direction, is  a simplification which allows to reduce the model towards a two-dimensional system. Since the strong magnetic field homogenizes the parallel dynamics, the perpendicular $E\times B$ velocity field is close to two-dimensional. This, in combination with the incompressibility of the $E\times B$ velocity and the assumption that the parallel dynamics is governed by a narrow spectrum peaked around a constant value $k_z$, allows to use the above set of equations for the scalars $\omega$ and $n$. 

The coupling term $c(\phi-n)$ permits the system to access to a saturated turbulent state even in the absence of external forcing. This is the main difference with the equations describing the two-dimensional mixing of a scalar in fluid turbulence. It is related to the presence of the parallel current-density, which couples the two-equations and gives rise to an electrostatic plasma instability leading to a saturated turbulent state in which the energy is drawn from the imposed mean plasma density profile. The collisionality of the ions and electrons plays a key role in the model. If the collisionality tends to a large value hence $c$ becomes small, the equations tend to a hydrodynamic 2D limit in which long-living vortices are observed. It was found in \cite{Bos2010-1} that the Lagrangian acceleration in this case showed a very intermittent behavior, reflected by probability density functions with heavy tails. For intermediate values of $c$ the flow is called quasi-adiabatic. The PDFs of the acceleration in this regime tend to exponential distributions.

One remaining open question is whether this intermittent behavior is a Reynolds number effect. Indeed in three-dimensional Navier-Stokes turbulence \cite{Vedula1999,Voth2002} the flatness of the acceleration PDF increases as a function of the Reynolds number for the Reynolds numbers currently available. In the present investigation this Reynolds-number dependence is analyzed by exploiting the results of a set of direct numerical simulations of the Hasegawa-Wakatani model for varying Reynolds number.

Another issue is the relation between the time-correlation of the norm of the acceleration and the manifestation of intermittency as proposed by Mordant \etal~ \cite{Mordant2002}. The present study will allow to assess this relation for the different regimes.

\begin{figure}
\centering
\setlength{\unitlength}{.5\textwidth}
\includegraphics[width=1.\unitlength,angle=0]{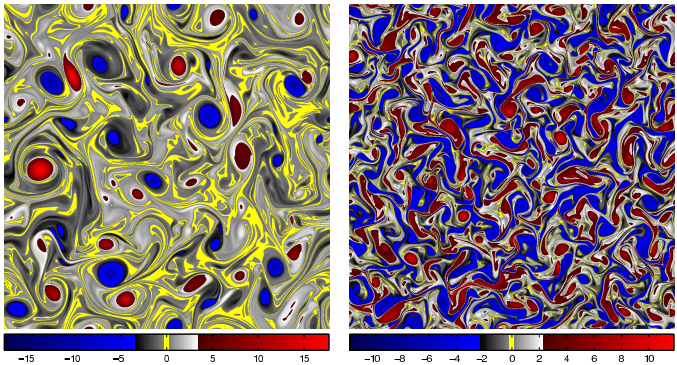}
\caption{Visualisations of the vorticity field for two different values of the adiabaticity. Left: $c=0.01$, right: $c=0.7$.
\label{visu} }     
\end{figure}

Equations (\ref{hw1},\ref{hw2}) were solved in a double-periodic
domain of size $64^2$ using a fully dealiased pseudo-spectral method
at a resolution of $1024^2$ gridpoints, starting from Gaussian random
initial conditions. In the saturated, fully developed turbulent flow $10^4$ particles were injected, equally spaced, and their velocity and acceleration were monitored during a large number of large-scale turn-over times {($\sim 400 T_e$). The eddy turn-over time $T_e$ obtained in the different regimes, defined as $1/\sqrt{\mathcal W}$ where $\mathcal W$ is the RMS vorticity, is of the same order of magnitude, $\sim 0.4$.} Details on the simulations of equations (\ref{hw1},\ref{hw2}) can be found in [\onlinecite{Bos2010-1}] and on the Lagrangian part of the study in [\onlinecite{Kadoch2008}] in which a similar investigation was performed for Navier-Stokes turbulence. The adiabaticity is varied between $c=0.01$ and $c=2$, to obtain different flow regimes. Visualizations of the vorticity-field for two flow regimes are shown in Figure \ref{visu}. 

\begin{figure}
\centering
\setlength{\unitlength}{0.5\textwidth}
\includegraphics[width=.7\unitlength,angle=0]{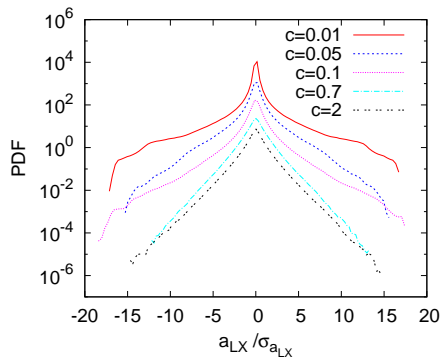}
\caption[]{PDFs of the Lagrangian acceleration ($x$-component) for different values of $c$. PDFs are normalized by $\sigma_{a_{LX}}$, the RMS value of the acceleration. Graphs for different $c$ are shifted vertically for clarity.} 
\label{pdfa}     
\end{figure}

\begin{figure}
\centering
\setlength{\unitlength}{.3\textwidth}
\includegraphics[width=1.\unitlength,angle=0]{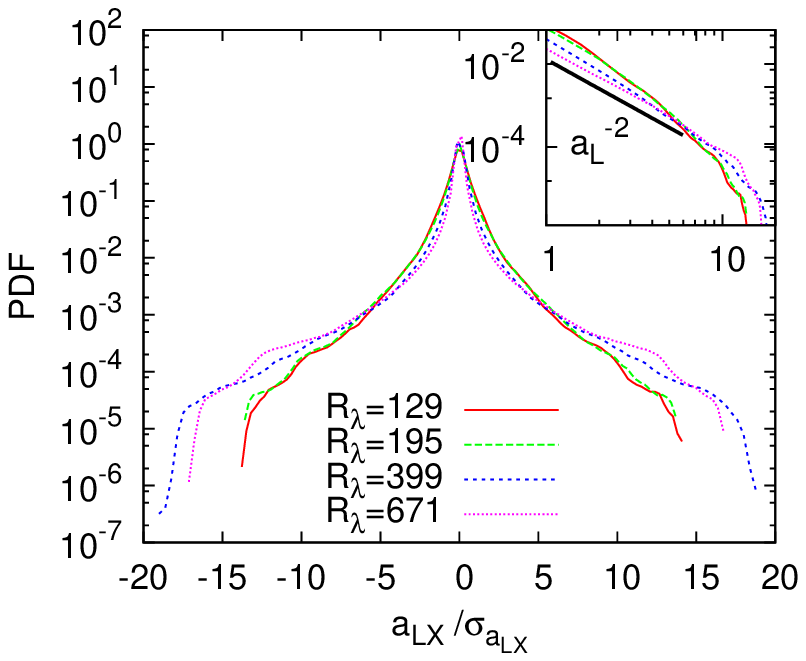}\\
\includegraphics[width=1.\unitlength,angle=0]{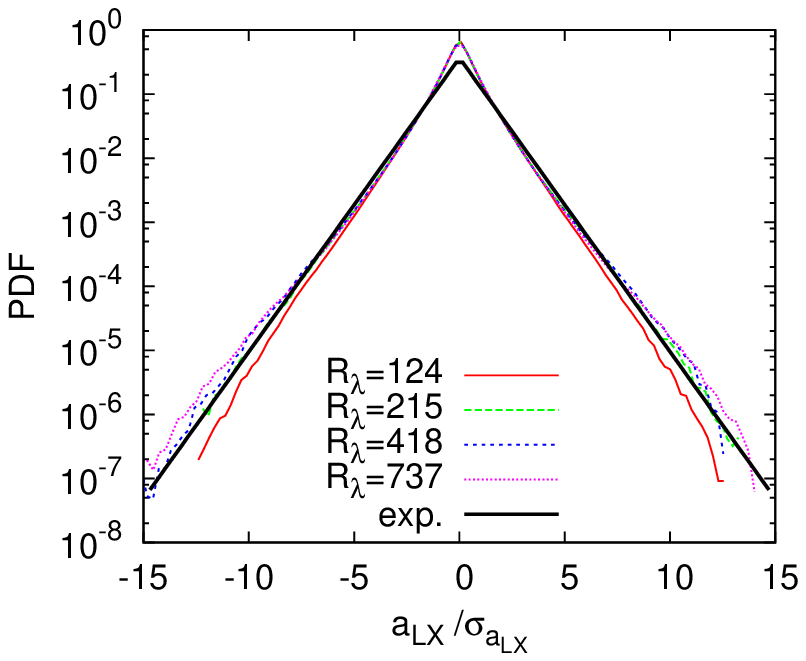}
\caption{
\label{pdfaRe} PDFs of Lagrangian acceleration at different Reynolds numbers for  $c=0.01$ (top) and $c=0.7$ (bottom). In the inset of the top figure, the power-law behavior of the tails of the PDF is demonstrated in double-logarithmic representation.}    
\end{figure}

In Figure \ref{pdfa}, the PDFs of the Lagrangian acceleration are shown for different values of $c$. It is observed that the PDF evolves from a close to exponential shape for large $c$ to an algebraic shape for $c=0.01$. To check if this is merely an effect of the Reynolds number, we performed simulations at different Reynolds numbers, which is here defined as $R_\lambda=\lambda~\mathcal U/\nu$, with $\mathcal U$ the RMS velocity and $\lambda=\mathcal U/\mathcal W$, an intrinsic scale of the turbulence. 
 This Reynolds number was varied by a factor $6$. The Prandtl number was kept unity. The results are shown in Figure \ref{pdfaRe}, where it is observed that the Reynolds number only slightly influences the shape of the PDFs.
Therefore we need  to find an alternative explanation for the difference in shapes of the accelerations for the two different flow regimes. The exponential distributions can be explained as follows: it was argued in [\onlinecite{Holzer1993}] that an exponential distribution for the pressure gradient PDF 
can be obtained from random Gaussian (non-intermittent) velocity fields by simply solving a Poisson-equation to obtain the pressure and subsequently computing the gradient, without considering the nonlinear dynamics of the Navier-Stokes equations. It can be seen from equation (\ref{eqaL}) that the pressure gradient is directly related to the Lagrangian acceleration. The shape of the PDFs for the cases for moderate and large $c$ simply shows that the flow is not intermittent from a Lagrangian point of view, but governed by a {Gaussian-like diffusion} process.

More puzzling are the algebraic tails, found for small $c$. In the
inset of Figure \ref{pdfaRe} we show that the tails show a close to algebraic behavior of the form $p(a)\sim 1/a^{\beta}$ with $\beta$ of the order $2$. It is interesting to note that the shape of the PDFs obtained in the hydrodynamic case closely ressembles the results obtained for point-vortices. Indeed, in reference [\onlinecite{Rast2009}] the point vortex model, introduced by Onsager \cite{Onsager1949} and Townsend \cite{Townsend1951}, was used to study the influence of point-vortices on the Lagrangian acceleration of passive tracers. In their work the acceleration PDF was to leading order given by $p(a)\sim 1/a^{5/3}$. In this light the results for the quasi-hydrodynamic flow seem to be at least partially explained by the presence of vortical structures as observed in Figure \ref{visu}. The exponent of the power-law tails of the acceleration PDF is close to the value $-5/3$ as in the point-vortex study. Even better agreement might be obtained by comparing with vortex-interaction models using vortices with a finite extension \cite{Wilczek2008-2}  (such as the Burger's vortex). 

\begin{figure}
\centering
\setlength{\unitlength}{0.5\textwidth}
\includegraphics[width=.7\unitlength,angle=0]{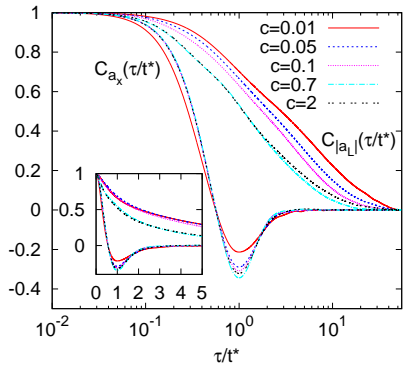}~
\caption[]{Autocorrelations of the acceleration $C_{a_{Lx}}$ and autocorrelation of the norm  of the acceleration $C_{|a_L|}$ for the five cases. {Inset: Autocorrelations of the acceleration in lin-lin.} For each curve time is normalized by $t^*$, the time at which the correlation component reaches its minimum value.}
\label{Ca}     
\end{figure}

It remains to be explained why this is not the case for the
 quasi-adiabatic case. As observed in Figure \ref{visu}, in this case
 the drift-waves also seem to organize into vortical
 structures. However the life-time of these structures is shorter
 \cite{Koniges1992}. The parallel dynamics are thus responsible for
 the change in life-time of the vortices. For higher adiabaticity,
 electrostatic fluctuations are rapidly smoothed out through the
 parallel current. Vortices do then not exist long enough to influence
 the acceleration statistics intermittently. In this sense the
 long-time correlations seem to be essential to obtain the algebraic
 tails in the acceleration PDF.  The centripetal component of the
 acceleration is constant in a purely circular orbit, and this is
 captured by the autocorrelation of the norm of the acceleration,
 which can therefore be directly related to the lifetime of the
 vortical structures. This is checked in Figure \ref{Ca}. For all
 curves time is  normalized by the time at which the autocorrelation
 of the $x$-component of the acceleration is minimum. This time-scale
 can be qualitatively related to the time-scale of the average
 circular motion of fluid-particles. 
 The autocorrelation of the acceleration components displays a behavior similar to what is observed in three-dimensional Navier-Stokes turbulence, with a rapid decrease and a negative dip. This dip becomes less pronounced for lower values of $c$. It is observed that in the cases in which a closer to exponential decay of the acceleration PDFs is observed, the auto-correlation of the norm decorrelates faster than in the cases in which the PDFs are algebraically decaying. Indeed the time correlations of the norm become longer for small adiabaticities. This constitutes a proof of the direct relation between the time-correlation of the norm of the acceleration and Lagrangian intermittency as proposed in \cite{Mordant2002}. A way to numerically check the assumption of the role played by time-correlations of the norm of the acceleration within a point-vortex model would be to vary the lifetime of the vortices. If short enough lifetimes are imposed, exponential tails are probably obtained. 

The main conclusion of the present work is that the electrostatic turbulence studied here  is not intermittent once the adiabaticity is large enough. This corresponds to the case in which the parallel structures have a short enough wavelength (or high parallel wavenumber) and/or small collisionality. Intermittency due to electrostatic vortex structures is therefore expected to be stronger near the edge of fusion plasmas, where the collisionality becomes more important. 

In the present study the transition between longliving structures and short lived wavy structures takes place somewhere in between $c=0.1$ and $c=0.7$. In reality the parallel spectrum is broadband and we assumed its peak around a certain frequency to obtain the simplified two-dimensional model. If the full three-dimensional model is considered, the dynamics will probably be a mixture between the different cases, dominated by a certain peak-wavenumber. Also the conclusions of this study relate to the dynamics captured within the present model, \emph{i.e.} homogeneous electrostatic turbulence fed by a strong plasma-density gradient. 

{For larger adiabaticity ($c>0.7$)}, which is expected to correspond to a situation further away from the edge or for colder plasmas, the statistics of this kind of turbulence are close to what would be expected from a Gaussian system. This study suggests that with respect to transport coefficients, microturbulence can be modeled by a Gaussian diffusion process with some additional rare point-vortices if the adiabaticity {is small enough ($c<0.7$)}.  This does not imply that plasma turbulence is not intermittent, only that its origin is not due to the mechanism contained in the present slab-geometry if the adiabaticity is large enough. It could be interesting to carry out a similar study in a more complete geometry, such as in the study by Holland \etal~ \cite{Holland2007}. In their work, dynamic regimes containing long-living vortices were observed, directly related to the large-scale zonal flows. However no fully developed turbulent state was considered. Studying the turbulent Lagrangian dynamics in such a geometry constitutes an interesting perspective. 

\emph{Acknowledgements.} Salah Neffaa is acknowledged for the validation of the numerical code and Diego Del Castillo Negrete, Sadruddin Benkadda and Shimpei Futatani for stimulating interaction.


\end{document}